# Retrieval of Particle Size Distribution of Polar Stratospheric Clouds Based on Wide-Angle Color and Polarization Analysis

Ugolnikov O.S.[1*], Kozelov B.V.[2], Pilgaev S.V.[2], Roldugin A.V.[2]

[1]Space Research Institute, Russian Academy of Sciences, Moscow Russia
[2]Polar Geophysical Institute, Apatity, Russia
*Corresponding author e-mail: ougolnikov@gmail.com

**Abstract.** The intensive polar stratospheric vortex in the Arctic that shifted to northern Europe in the winter of 2019-2020 caused low temperatures and the frequent occurrence of polar stratospheric clouds followed by a significant decrease in the total ozone content. Polarimetry and multi-color photometry using all-sky cameras in Lovozero (Murmansk Region, Russia, 68.0°N, 35.1°E) together with a new method of cloud field separation against the twilight background allowed for finding the light scattering characteristics related to the particle size distribution of polar stratospheric clouds. The results are compared with lidar and balloon experiments. The conditions of the appearance of visually bright polar stratospheric clouds of type I are discussed.

## 1. Introduction

Polar stratospheric clouds (PSC) were first observed in March 1870 in northern Scandinavia (Kassner, 1895). A few decades later, PSC were also noticed during the Antarctic winter in the Southern hemisphere (Arctowskiy, 1902). Clouds can remain bright during the twilight, which shows that their altitude is not less than 20 km. Sometimes, the fragments of the cloud had intensive color changing from red to blue. Due to this irisation effect, these clouds are also called "nacreous" or "mother-of-pearl". This can take place if the particles forming the cloud have similar sizes that significantly larger than the optical wavelength.

The chemical nature of the clouds became clear in the late 20th century. PSC particles are nucleated on stratospheric aerosol consisting of a $H_2SO_4$ solution (Rosen, 1971). As the temperature falls below 195 K, nitric acid $HNO_3$ condenses on these particles (Crutzen and Arnold, 1986, Toon *et al.*, 1986). Depending on the physical conditions and cooling rate, these can form liquid supercooled ternary solution $H_2O/H_2SO_4/HNO_3$ (STS, Molina *et al.*, 1993, Carslaw *et al.*, 1994) or solid particles of nitric acid trihydrate (NAT) (Tabazadeh *et al.*, 1994). These particles form PSC of types Ib and Ia, respectively.

Solid NAT particles have a size of more than 1 μm, that is, several times larger than liquid STS particles (Voigt *et al.*, 2000); however, the numeric density of solid particles is 30-100 times lower. Taking into account that the optical scattering coefficient increases with the mean particle radius $r_0$ very slowly for $r_0$ from 0.2 to 2 μm, liquid particles have a stronger mass-equivalent optical effect compared with solid particles. While STS particles are spherical, solid NAT particles are not, and light scattering is not exactly described by the Mie theory. The basic observational effect of non-sphericity is the depolarization of backscattering fixed by lidar sounding (Browell *et al.*, 1990).

If the temperature falls below the ice frost level (about 185 K in lower stratosphere conditions), the nucleation of water ice onto a particle begins. Ice crystals can reach a size of more than 10 μm forming bright and color-variable clouds of type II (Poole and McCormick, 1988). They can be seen as bright fragments against the background of type I PSC.

PSC have been an object of special interest during the last few decades due to heterogeneous reactions releasing active chlorine-containing molecules and radicals that destroy stratospheric ozone (Toon *et al*., 1986; Solomon *et al*., 1986; Solomon, 1990). PSC particles are studied directly from balloons (Deshler *et al*., 2000). Lidar measurements are effective to distinguish between different types of PSC particles using a cross-polarization scheme (Browell *et al*., 1990) and also to find the particle size distribution (Jumelet *et al*., 2009). This advantage makes an effective combination of backscattering analysis by combining it with other measurements (Deshler *et al*., 2000) and the lidar sounding of PSC from space (Noel *et al*., 2008).

The ground-based passive optical sounding of PSC particles seems difficult owing to the rare occurrence of bright clouds with a high S/N ratio against the twilight background, weather restrictions, and a low number of observational sites in high latitudes. However, anomalies of the Arctic stratosphere polar vortex during the winter of 2019-2020 led to a number of bright events over northern Russia. In this paper, the authors suggest the method of PSC particle size distribution retrieval based on all-sky measurements and compare the results with existing data.

**2. Observations**

Polar stratospheric clouds were detected during the winter of 2019-2020 by all-sky cameras for regular sky surveys in the Lovozero station (68.0°N, 35.1°E) of the Polar Geophysical Institute. This work is based on the measurements of color and polarization cameras. Both devices had a field size diameter of 180°; the sky area with zenith angles of up to 70° was analyzed. The cameras worked during both the twilight and night; star images were used to fix the camera position and the flat field. For PSC analysis, solar zenith angles $z$ up to 94° were considered. The lower limit of the $z$ interval was 89.5° or more if it was restricted by local polar night conditions.

The color camera had an RGB detector and an IR-cutting filter with a threshold wavelength of 680 nm. The B, G, and R instrumental bands had effective wavelengths of 460, 530, and 595 nm, respectively; full width at half maximum (FWHM) was close to 90 nm for all bands.

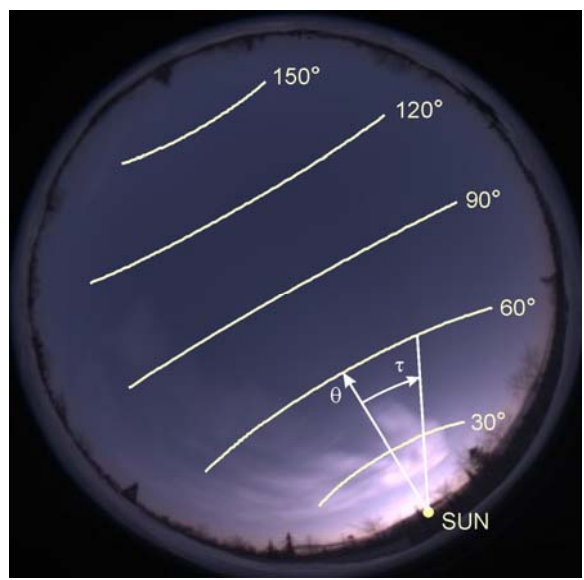

Another camera with a rotating polarization filter worked in the spectral band with an effective wavelength of 530 nm and FWHM of 70 nm, close to the G band of the color camera. The procedure of polarization measurements was described by Ugolnikov and Kozelov (2016). The sky image diameter was about 500 px for both cameras.

Bright PSC were observed in clear troposphere conditions on the evening of November 27, 2019, the morning and evening of January 1 and 27, 2020. Figure 1 shows the image of PSC on the evening of January 27, the solar zenith angle is 89.7°. A wide amorphous structure without special bright spots was the general property of the PSC field observed this winter, which is typical for PSC of type I. However, separate PSC of type II are possible. This depends on stratospheric temperatures, which can be checked by satellite data.

*Figure 1. All-sky image with polar stratospheric clouds (evening January 27, 2020) with coordinate system definitions and lines of constant scattering angle.*

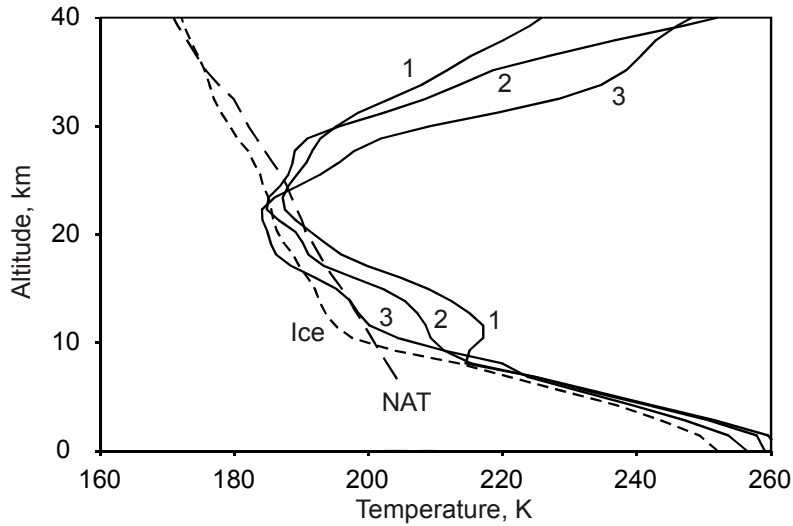

*Figure 2. EOS Aura/MLS temperature profiles above the observation point for PSC occurrence dates (1 - November 27, 2019, 2 - January 1, 2020, 3 - January 27, 2020). NAT and ice frost points profiles for January 27 are shown.*

Figure 2 shows the vertical profiles of temperature by nearby EOS Aura/MLS scans (EOS Team, 2011) of the observation dates; the profiles of NAT and ice frost temperatures based on MLS data for January 27 are also shown. It can be seen that atmospheric conditions met the formation of STS and NAT particles in all cases, and ice particles were also fragmentally possible. Type II clouds can reveal themselves as bright regions of variable color significantly contributing to the errors of wide-field color analysis of type I clouds that must be checked during the analysis.

**3. Cloud field separation**

Fixing the clouds scattering field against the twilight background and measuring its observational characteristics (polarization or the intensity ratio in different color bands) are the most difficult problems of cloud study by all-sky analysis. Clouds do not have sharp borders, and the twilight background is also spatially variable. Its brightness significantly changes from date to date, and one can not simply subtract the background measured during another twilight without PSC.

Solving this problem for the polarization study of noctilucent clouds (NLC), Ugolnikov *et al.* (2016) excluded low-degree harmonics of twilight background variations along the line of the constant scattering angle. This procedure was followed by the orthogonalization of residuals with cloud-free sky variations measured during the reference twilight. It worked well for the case of bright NLC (Ugolnikov and Maslov, 2019), but here are no reliable reference twilight data with the same atmospheric conditions, solar azimuth, and scattering geometry.

Another possible approach used for the color study of NLC (Ugolnikov *et al.*, 2017) was Fourier analysis along sky almucantars followed by the exclusion of lower harmonics. It was also appropriate for bright NLC with a fine wave structure, but the complete exclusion of the background requires a significant loss of the S/N ratio, which can be a larger problem for the amorphous field of type I PSC. It should also be noted that the brightness of PSC scattering during the observations rarely exceeded 50% of the twilight background, usually being a small percentage of this value. In this case, the authors built a new method of background subtraction applicable for color and polarization data. The results are to be compared with the methods for the both color and polarization analysis of NLC listed above (see Appendix).

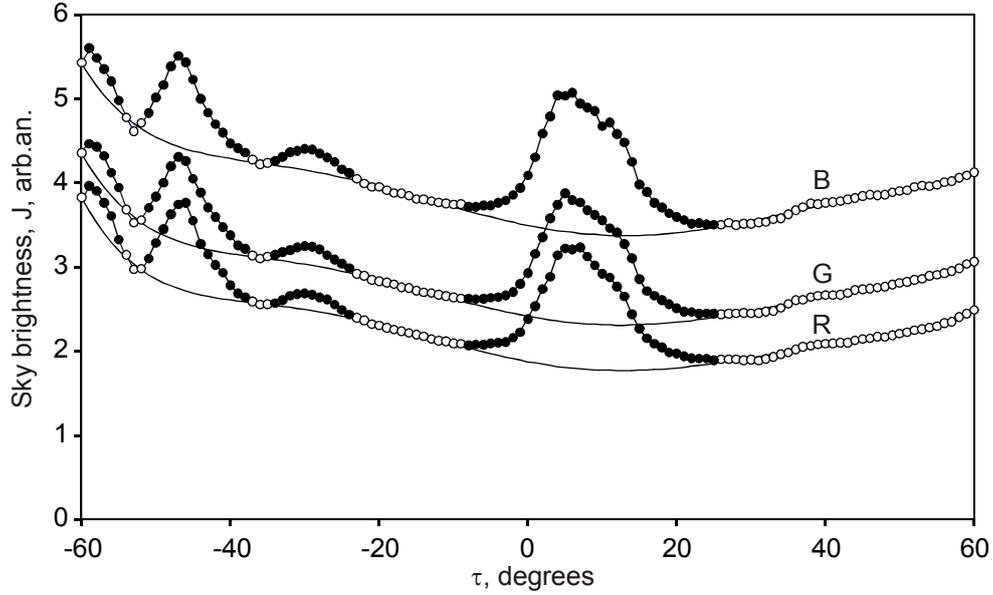

*Figure 3. Example of scan of the sky with θ=45° for January 27.
Open circles are associated with background, black dots are related with PSC.*

Following Ugolnikov *et al.* (2016), we define a sky polar coordinate system, where one pole coincides with the Sun (see Figure 1). Here θ is equal to a single scattering angle of the atmospheric element seen in this sky point (for stratospheric trajectories, the refraction angle is less than 1° and does not significantly change the scattering angle). The lines of equal θ are shown in Figure 1; τ is another coordinate changing along these lines. The observational data is integrated with 1° resolution for the color camera and 2° resolution for the polarization camera.

Figure 3 shows the scans of the sky along the line with fixed θ=45° for the image shown in Figure 1 (open and black dots). Taking the profile in color G, $J_G(\tau)$ as the basic, there are $N_0$ experimental points ($N_0$=121 for the case in the figure). At the initial stage, it is necessary to find the *M*-degree polynomial fit of this profile $I_{G0}(\tau)$ and calculate the deviation:

$$\sigma_0^2 = \frac{\sum_{i=1}^{N_0}(J_G(\tau_i) - I_{G0}(\tau_i))^2}{N_0 - M - 1}.$$

(1)

Then, one can find the point $\tau_{i1}$ with the maximal positive residual $J_G(\tau_{i1}) - I_{G0}(\tau_{i1})$. This sky point is related to PSC and dropped out from the following clear sky analysis. In the next stage, there are $N_1=N_0-1$ points, and it is possible to calculate the new approximate profile $I_{G1}(\tau)$ and deviation:

$$\sigma_1^2 = \frac{\sum_{i \neq i1}(J_G(\tau_i) - I_{G1}(\tau_i))^2}{N_0 - M - 2}.$$

(2)

After that, one can find the point $\tau_{i2} \neq \tau_{i1}$ with the maximal positive residual $J_G(\tau_{i2}) - I_{G1}(\tau_{i2})$ and repeat this process again. Since the excluded point is characterized by a large residual, the value of σ will decrease for each next step of the procedure. In the case of a noisy background without PSC, it is expected that $\sigma_K$ will not decrease faster than the number of points that remain, $N_K=N_0-K$. However, if one excludes physically brighter points, σ can decrease more rapidly. Thus, there is the following criterion for process termination:

$$\frac{\sigma_K}{N_K} = \frac{\sqrt{\sum_{i \neq i1,i2,...iK}(J_G(\tau_i) - I_{GK}(\tau_i))^2}}{(N_0 - K)\sqrt{N_0 - K - M - 1}} = \min. \tag{3}$$

The result of this process with $M=6$ is shown in Figure 3. All the points associated with PSC are black dots; the background points that remained in the analysis are shown by open circles. The procedure is done only for the G profile; the final background approximations for the B and R profiles $I_{BK}(\tau)$ and $I_{RK}(\tau)$ are based on the same sample of $\tau_i$ (open circles) which was finally chosen for the G profile.

During the polarization analysis, color profiles are substituted for the profiles of the Stokes components for the sky background $J_{1,2}(\tau)$ in the coordinate system shown in Figure 1. The basic profile for the analysis described above is the intensity profile $J_1(\tau)$. Second Stokes component $J_2(\tau)$ is positive if polarization is directed along the line ($\theta$=const) and negative if it is directed along the line ($\tau$=const) or parallel to the scattering plane. Having found the background points $\tau_i$, from analysis of intensity $J_1(\tau)$, we approximate the background profiles $I_{1,2K}(\tau)$.

It should be added that this procedure does not involve finding the lower envelope curve. In that case, one would not eliminate the background noise from the PSC signal. The open circles in Figure 3 can be above and below the profile (thin line). In rare cases, if PSC are expanded along the whole line ($\theta$=const) in the sky, one can lose part of the PSC signal but this effect will be similar for all channels and thus will not affect the color and polarization estimation. The influence of the sky background after this procedure is still possible and will be taken into account below.

Let us define the cloud signal $j(\tau)=J(\tau)-I_K(\tau)$ and combine the black-dotted data in the neighbor points $\tau$ for which $j_B(\tau)>0$, $j_G(\tau)>0$ and $j_R(\tau)>0$ for the color data and $j_1(\tau)>0$ for the polarization data. For example, the profile in Figure 3 contains four such "spots". For each spot, the following geometric characteristics are determined: the solar zenith angle in the observational point $z$, the local solar zenith distance $z_L$ observed from the cloud, the scattering angle $\theta$, the mean $\tau$ angle, and the zenith angle $Z(z,\theta,\tau)$. We also find the physical characteristics: the color ratios $C_R=j_R/j_B$ and $C_G=j_G/j_B$ or the polarization degree $p=j_2/j_1$. Another important parameter is the ratio of cloud and background integrated signals $\eta=j_G/I_{GK}$ (color analysis) or $\eta=j_1/I_{1K}$ (polarization analysis). Spots with $\eta>0.02$ are considered.

## 4. Polarization analysis

The polarization of PSC light scattering is the function of the angle $\theta$. There is a number of measurements of spots with a definite angle $\theta$. The simplest way is to find the average value as it was done for NLC by Ugolnikov *et al.* (2016), and Ugolnikov and Maslov (2019). However, these data can be also used to check the influence of the incompletely eliminated sky background. It can be assumed that polarization is measured for the sum of the cloud field $j_1(\tau)$ and background admixture $\alpha I(\tau)$, where $\alpha$ is a small parameter. If the polarization of PSC scattering is $p_C$ and background polarization is $p_S$, then the measured polarization will be equal to:

$$p = \frac{p_C j + p_S \alpha I}{j + \alpha I} \approx p_C + (p_S - p_C)\frac{\alpha I}{j} = p_C + (p_S - p_C)\frac{\alpha}{\eta}. \tag{4}$$

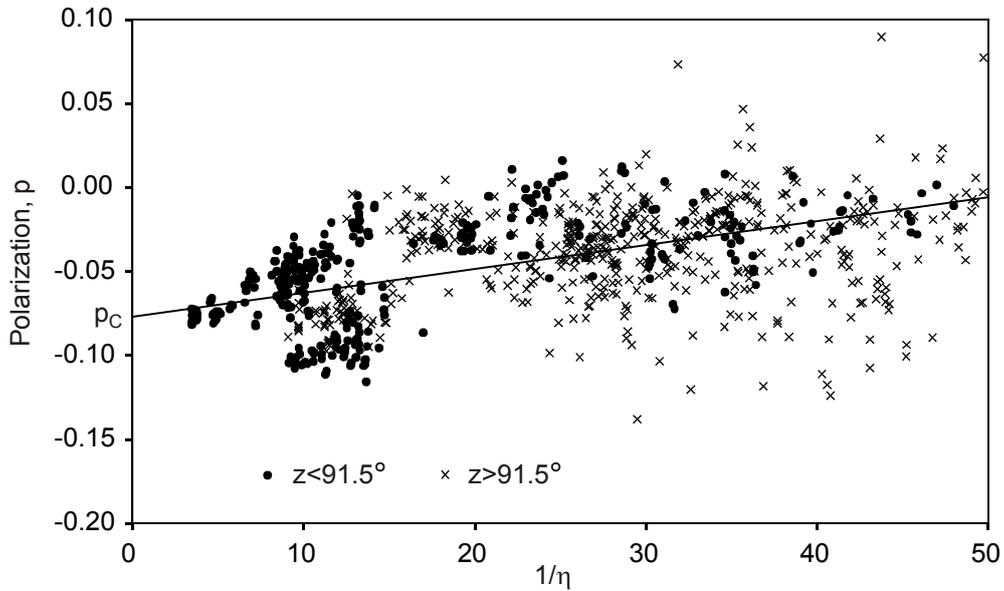

*Figure 4. Procedure of determination of polarization of PSC light scattering (θ=40°).*

Here, it is assumed that $\alpha I \ll j$. Variations of sky polarization $p_S$ along the line of constant θ are less than the difference $p_S - p_C$, and one can build the linear relation of the measured values of $p$ and ($1/\eta$). An example of this dependence for θ=40° is shown in Figure 4. The observational points are strongly shuttered especially in the right part of the figure (faint spots with small η), which is caused by errors with the polarization measurements and variations of the parameter α. The slope of dependency is natural since the background consists basically of Rayleigh scattering with a positive value $p_S$. Taking the weight of each measurement equal to its value of η, one can run the least square procedure and find $p_C(\theta)$.

Another remarkable thing that one can see in Figure 4 is independency of measured PSC polarization on the twilight stage - agreement between the measurements at different solar zenith angles $z$ shown by dots and crosses in figure. We can consider this as the confirmation of correct "spots" separation and small influence of scattering of background sky emission on PSC particles (multiple scattering on PSC). Its polarization differs from single scattering, and if its contribution was higher, then we could expect polarization changes during the twilight, as PSC is immersed into the shadow of the Earth.

Procedure of $p_C$ determination works well if there are a number of measurements of bright spots with a high value of η. For PSC described here, this is true for scattering angles from 30° to 60°; accuracy sharply decreases for larger angles. Figure 5 shows the dependencies of $p_C(\theta)$ for two twilights on January 27, 2020. The basic property is a negative sign of the second Stokes component meaning that polarization is directed parallel to the scattering plane. It is the basic difference from scattering on small particles observed in NLC where the value of $p_C$ is positive and can reach unity near θ=90° (see Appendix). Reverse polarization at these angles is the usual property of spherical particles with radii of above 0.2 μm; theoretical dependencies for some particle ensembles described below are also shown in the figure. However, it can be seen that the dependence of polarization on the mean particle size is not monotonous and it is not possible to estimate the size based on polarization only as it was done for NLC particles. Color data must be added; this analysis is done in the next chapter of the work.

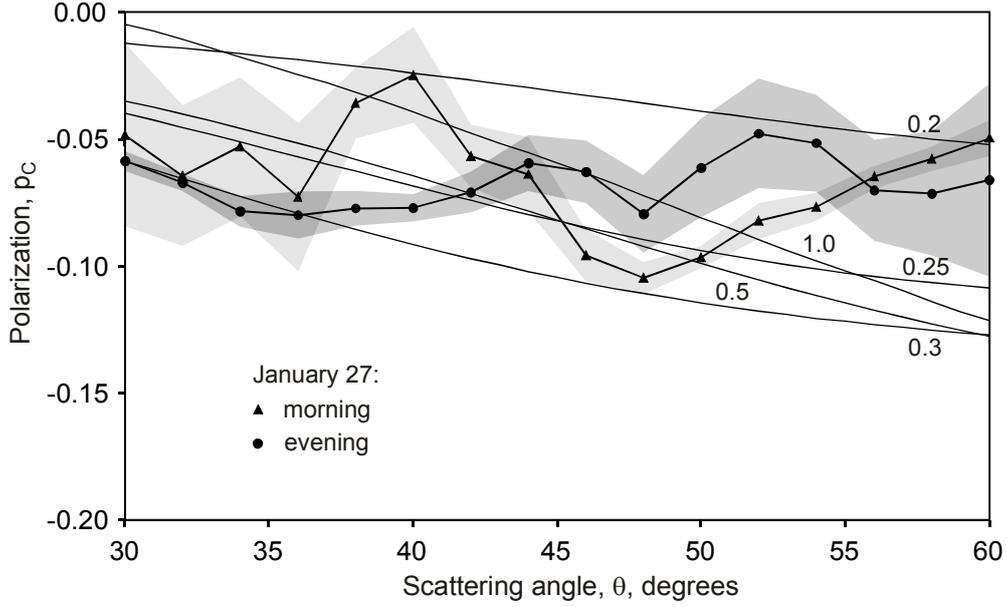

*Figure 5. Polarization of PSC light scattering compared with theoretical calculations for refractive index 1.47 and lognormal size distribution with $\sigma_D$=1.45. Median radii in μm are denoted.*

## 5. Color analysis

Multi-wavelength effects in PSC are considered in a way similar to NLC analysis (Ugolnikov *et al.*, 2017). The basic difference is that PSC particles are significantly larger than NLC particles. The Rayleigh-Gans approximation is applicable for NLC; in this case, the color ratio of Mie scattering functions $S_{G,R}(\theta)/S_B(\theta)$ is a linear function of $\cos\theta$, the coefficient is simply related to the mean particle size. For the particle radius above 0.1 μm, this dependency becomes more complicated. The particle size distribution is assumed to be lognormal:

$$f(r) = \frac{1}{\sqrt{2\pi}\,\zeta\, r}\exp\left(\frac{-\ln^2(r/r_0)}{2\zeta^2}\right); \quad \zeta = \ln\sigma_D. \tag{5}$$

Here $r_0$ is the median particle radius, $\sigma_D$ is the distribution width. Figure 6 shows the ratio $S_R(\theta)/S_B(\theta)$ for $\sigma_D$=1.45 and the refractive index $m$=1.47. It reflects the common properties of Mie scattering for the values of $\sigma$ and $m$ typical for type I PSC. The median particle radii $r_0$ in μm are denoted in the figure. The dependencies $S_G(\theta)/S_B(\theta)$ have the same properties as $S_R(\theta)/S_B(\theta)$.

It can be seen that the color ratio changes with the scattering angle in the most complicated manner near $\theta$=0° and $\theta$=180°. However, there is an interval from $\theta$=40° to about 110-120°, where the color ratio angular dependency is smooth and, in some cases, close to linear by $\theta$. Dropping out the color data for $\theta$<40°, one can measure the mean color gradient of PSC and compare it with theoretical data for different particle ensembles. It is not necessary to set the upper limit for this interval since most observational data is obtained for $\theta$<90°, just rarely reaching 110-120°. Since the dependencies shown in Figure 6 are not exactly linear, the mean color gradient can be defined as $W_{G,R}(m, r_0, \sigma_D)$, assuming the approximated relation

$$\frac{S_{G,R}}{S_B}(\theta) = \frac{S_{G,R}}{S_B}(\theta_0)\cdot(1+W_{G,R}(\theta-\theta_0)), \tag{6}$$

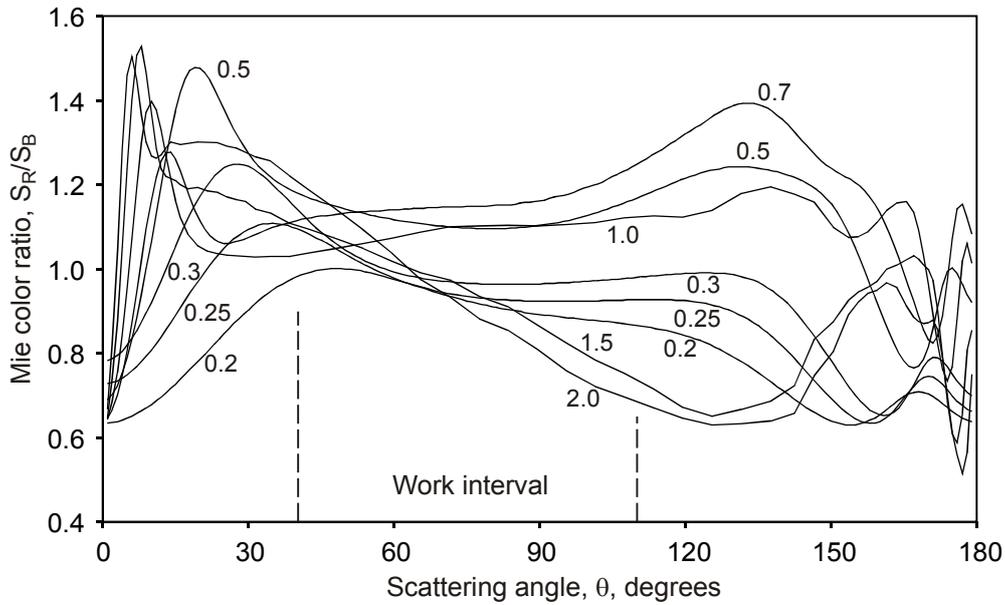

*Figure 6. Color ratio of Mie scattering for refractive index m=1.47
and lognormal size distributions with $\sigma_D$=1.45. Median radii in μm are denoted.*

where $\theta_0$ is equal to 60°. The value of $W_{G,R}$ is found from the theoretical profiles of $S_{G,R}/S_B$ by the least-squares method while each point by θ with step 1° is taken with a weight equal to the sum of the relative brightness values η of all registered "spots" at this scattering angle. The contribution of points with 40°≤θ≤60° is principal.

The color ratios $C_R=j_R/j_B$ measured for PSC on the evening of January 27 are shown in Figure 7; the data with η>0.05 is plotted. It can be seen that clouds get blue ($C_R$ decreases) far from the Sun that they are. The absolute color ratio measured by the camera does not coincide with the natural value since the camera sensitivity and solar spectrum in the B, G, and R spectral band profiles are different. The observed color gradient is stronger than $W_R$ since there are other factors influencing the color of PSC.

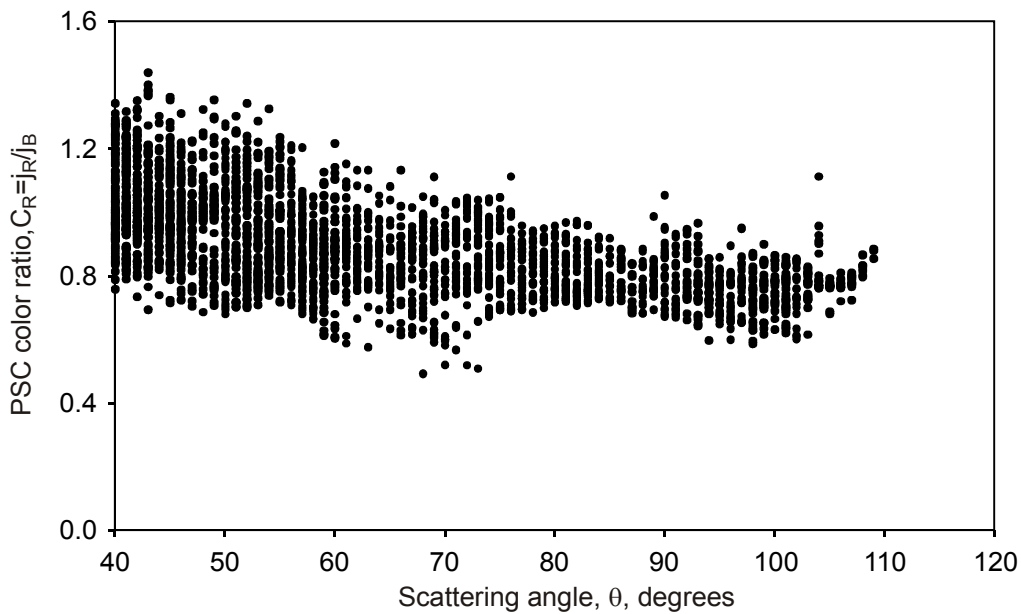

*Figure 7. Color ratios of spots in the evening of January 27.*

First, the conditions of solar illumination of PSC change during the twilight due to Rayleigh and aerosol extinction and ozone Chappuis absorption of solar emission before scattering. This factor is defined by the value of the local solar zenith angle $z_L$ observed from the cloud. Based on the mean altitude of PSC (20 km, based on the temperature profile for January 27 in Figure 2), the difference of the solar zenith angle in the cloud and observation point does not exceed 0.3°; however, it is taken into account. The color ratios $C_{G,R}$ are multiplied by $Q_{G,R0} \cdot (1+Q_{G,R}(z_L-z_{L0}))$, where $z_{L0}$ is the average solar zenith angle for the observation period, 92°.

Second, color is changed by extinction in the troposphere after scattering. Following Bouger's law, the color ratios $C_{G,R}$ are multiplied by $\exp(-T_{G,R}/\cos Z)$, where $T_{G,R}$ is the difference of troposphere optical depths in the G(R) and B channels. Finally, one can obtain the expression of the observed color ratio of PSC:

$$C_{G,R} = \frac{F_{G,R}}{F_B} \cdot \frac{S_{G,R}}{S_B}(\theta_0) \cdot (1+W_{G,R}(\theta-\theta_0)) \cdot Q_{G,R0}(1+Q_{G,R}(z_L-z_{L0})) \cdot \exp(-T_{G,R}/\cos Z). \quad (7)$$

Here $F_{B,G,R}$ is the multiplication of solar spectrum and camera sensitivity in the B, G, and R channels. Each factor described above changes the color ratio by not more than 10-20%, and the total color variations are not strong, as can be seen in Figure 7. In this case, one can combine all constants and use the linearization scheme in a way similar to (Ugolnikov *et al.*, 2017):

$$C_{G,R} = C_{G,R0} \cdot \left(1 + W_{G,R}(\theta-\theta_0) + Q_{G,R}(z_L-z_{L0}) + T_{G,R}\left(\frac{1}{\cos Z_0} - \frac{1}{\cos Z}\right) + \frac{A_{G,R}}{\eta}\right);$$

$$C_{G,R0} = \frac{F_{G,R}}{F_B} \cdot \frac{S_{G,R}}{S_B}(\theta_0) \cdot Q_{G,R0} \cdot \exp(-T_{G,R}/\cos Z_0). \quad (8)$$

Here $Z_0=45°$ (the mean zenith angle of bright PSC in analysis). The last term ($A_{G,R}/\eta$) is added to fix the possible influence of the incompletely eliminated sky background, in a way similar to polarization analysis. In fact, this term is found to be quite small. All the parameters $C_0$, $W$, $Q$, $T$, and $A$ can be found by the least-squares method; each point is taken with the weight equal to $\eta$. The results are the values of $W_G$ and $W_R$ that can be compared to theoretical calculations by the Mie theory and approximation (6).

**6. PSC size distribution retrieval**

The basic difference between polarization and color analysis is that the polarization $p$ is estimated directly and compared with theoretical calculations for different particle ensembles, while the color ratio of PSC scattering remains unknown, only its gradient $W$ by the scattering angle $\theta$ can be found. Polarization is measured in a short interval of $\theta$, its variations over this interval are not significantly stronger than errors for its measurements. In this case it is enough to determine the average $p_0$ value in the interval of $\theta$ from 30° to 60°. It is listed in Table 1 for all twilights with observed PSC.

The first thing that should be noted is that the values of $W$ are always negative, i.e. PSC become bluer the farther that they are from the Sun. This property is the opposite of NLC with small particles that get bluer the nearer that they are to the Sun in the sky (Ugolnikov *et al.*, 2017), but true for most parts of larger particle ensembles in Figure 6. The same can be said about the negative polarization of PSC light scattering.

| Date | $p_0$ | $W_G$, rad$^{-1}$ | $W_R$, rad$^{-1}$ | $s_G$ | $s_R$ |
|---|---|---|---|---|---|
| Nov 27, evening | 0.01 ± 0.17 | −0.13 ± 0.06 | −0.22 ± 0.11 | 0.13 | 0.23 |
| Jan 1, morning | no meas. | −0.19 ± 0.02 | −0.53 ± 0.06 | 0.07 | 0.17 |
| Jan 1, evening | −0.05 ± 0.12 | −0.04 ± 0.12 | −0.12 ± 0.22 | 0.08 | 0.18 |
| Jan 27, morning | −0.07 ± 0.02 | −0.05 ± 0.02 | −0.17 ± 0.05 | 0.07 | 0.18 |
| Jan 27, evening | −0.067 ± 0.010 | −0.104 ± 0.011 | −0.168 ± 0.016 | 0.06 | 0.10 |

*Table 1. Polarization and color characteristics measured for PSC in 2019-2020.*

Table 1 also contains the values of the standard deviation of the experimental color data from the expression (8), $s_G$, and $s_R$. It can be seen that they are not overly high, especially on January 27, 2020. It can also be added that all bright regions with η>0.3 have deviations of not more than 1.5·$s_{G,R}$. Thus, spots of type II PSC, if they ever were to appear, do not significantly distort the observed color structure at least on the day under consideration. The accuracy of $p_0$ and $W$ determination on January 27, especially in the evening, is sufficient to find the parameters of PSC size distribution.

The diagram "$p_0 − W_R$" is shown in Figure 8. The experimental results are shown together with theoretical calculations for the refractive index $m$=1.47 and size distribution widths $\sigma_D$=1.40, 1.45, and 1.50, the $r_0$ values are denoted. These parametric lines have a remarkable "treble clef" form, the observational dots lie just close to self-crossing point of the line for $\sigma_D$=1.45, which means that there are two possible median radii for these $m$ and $\sigma_D$. Multiple solutions are usual in the optical sounding of PSC as it was also noted in lidar experiments (Jumelet *et al*., 2009). Here, possible values of the median radius are close to 0.25 and 1.3 μm, which is typical for both types Ib and Ia PSC, respectively. However, the refractive index of PSC Ia particles is about 1.55, which marginally shifts the "treble clef", but a solution with a different $\sigma_D$ and $r_0$ can exist.

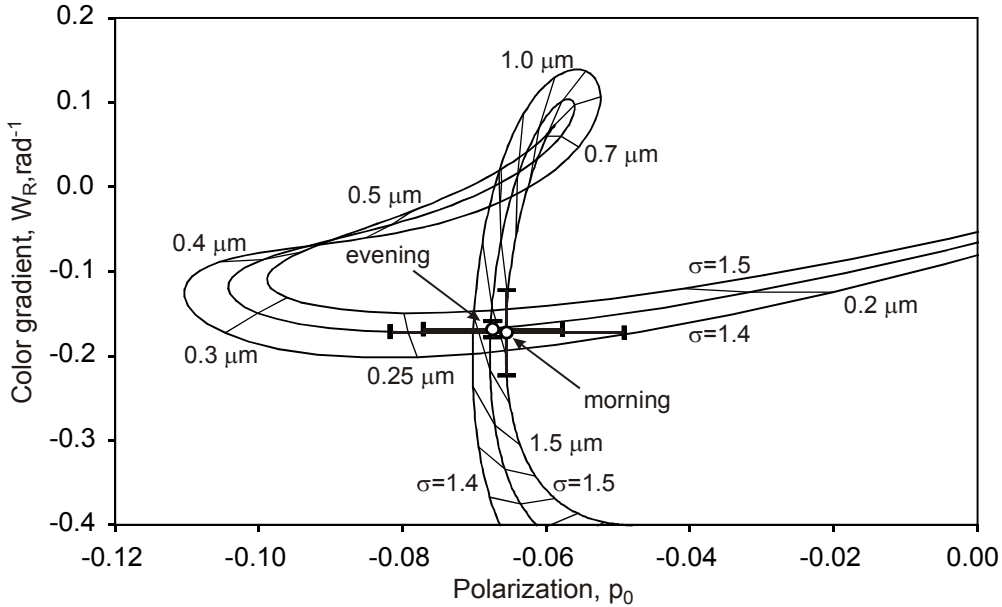

*Figure 8. Diagram "polarization - color" for measurements on January 27 and Mie scattering for refractive index m=1.47 and lognormal size distributions with $\sigma_D$=1.40, 1.45, and 1.50. Distribution widths and median radii are denoted.*

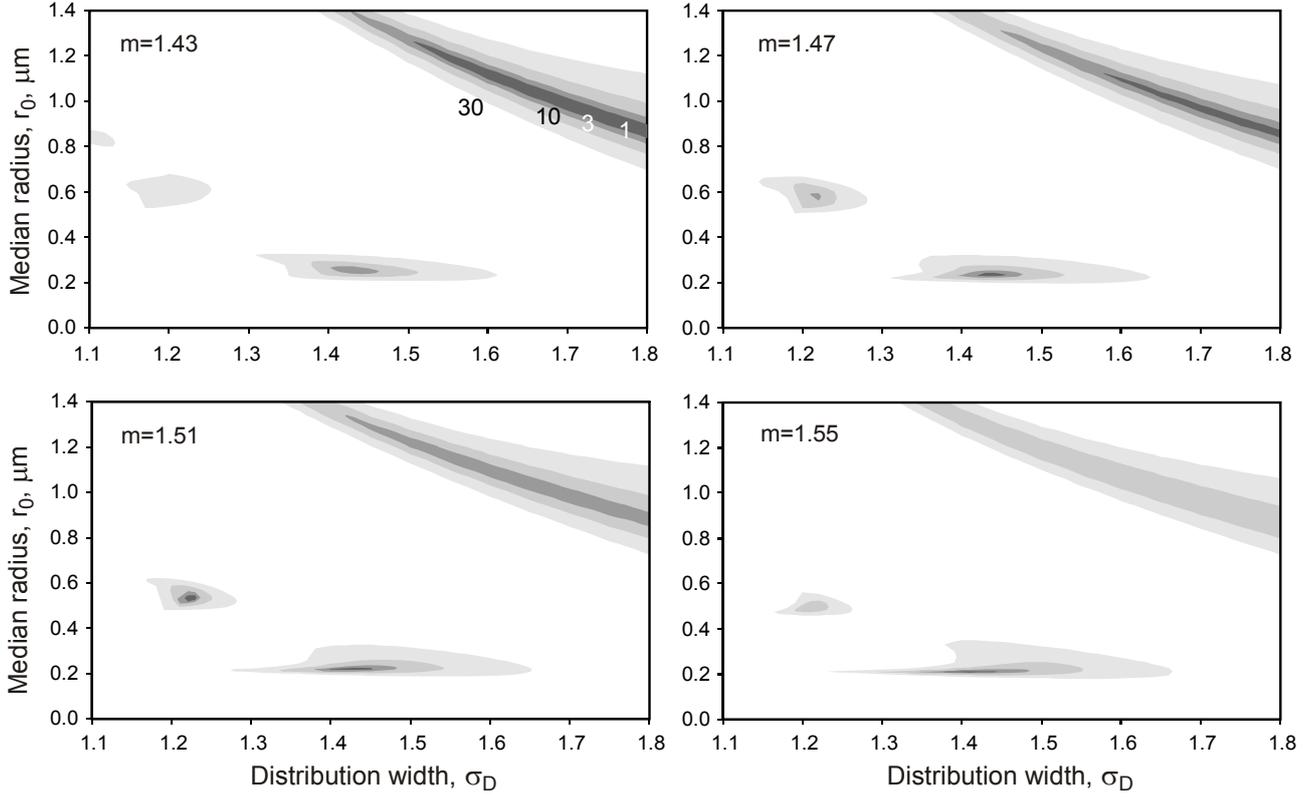

*Figure 9. Diagram of lognormal size distribution parameters retrieved from the observations in the evening of January 27. $\chi_3^2$ values are denoted in the upper left.*

To study this generally, one can take all the three measured parameters $p_0$, $W_G$, and $W_R$ and calculate the criterion of fitting the theoretical model:

$$\chi_3^2 = \left(\frac{p_0 - p_{0T}(m, r_0, \sigma_D)}{\sigma_p}\right)^2 + \left(\frac{W_G - W_{GT}(m, r_0, \sigma_D)}{\sigma_{WG}}\right)^2 + \left(\frac{W_R - W_{RT}(m, r_0, \sigma_D)}{\sigma_{WR}}\right)^2. \qquad (9)$$

Here the index "*T*" means the theoretical value, $\sigma_p$, $\sigma_{WG}$, $\sigma_{WR}$ are the errors of experimental parameters. One can calculate $\chi_3^2$ for different samples ($m$, $r_0$, $\sigma_D$) possible for PSC.

Maps of $\chi_3^2$ distribution in the "$\sigma_D - r_0$" diagram are presented in Figure 9 for different refractive indices on the evening of January 27. The values $\chi_3^2$ are denoted in the figure. The first thing that should be noted is the weak dependence of the picture on the refractive index. This means that particle size retrieval can be carried out even if the index is not exactly known.

Three types of possible PSC size distribution are clearly seen. The polarization and color data are not enough to distinguish between them and to be able to confidently choose the correct type. The same situation can take place in the lidar sounding of PSC (Jumelet *et al.*, 2009). It is necessary to consider the physical reliability to find what type is most possible.

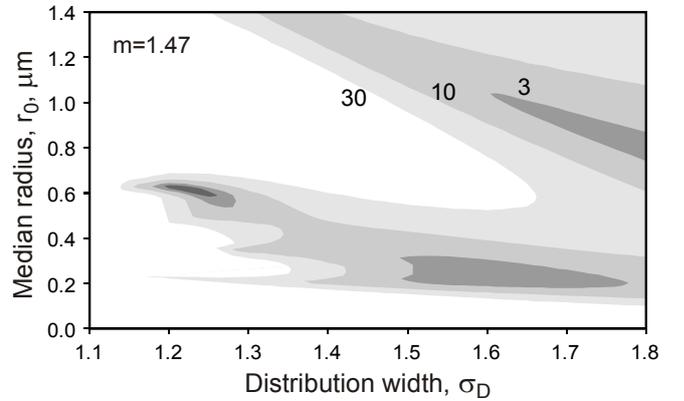

*Figure 10. Diagram of lognormal size distribution parameters retrieved from the observations in the morning of January 27, m=1.47. $\chi_3^2$ values are denoted.*

The first type of size distribution with the largest particles was found above by the analysis of Figure 8. The median particle radius is about 1 μm or more, which is typical for type Ia. Different distribution widths are possible, a minimum of $\chi_3^2$ is reached for wider distributions with $\sigma_D \geq 1.6$. In this case, the possible median radius decreases. It should be noted that PSC Ia particles are solid and non-spherical, and these estimations based on Mie theory can be considered as just an approximation. This type of size distribution seems to be less probable for bright clouds, since clouds of type Ia have less visual brightness compared with type Ib, as it was discussed in the beginning of the paper. For refractive index 1.55, $\chi_3^2$ was above 3 for the evening of January 27.

Second, the intermediate type is narrow distribution with a medium size. Minimal $\chi_3^2$ is reached for the refractive index 1.51, the best-fit median radius $r_0$ is equal to 0.53±0.02 μm, the distribution width is $\sigma_D$=1.22±0.02. The median radius increases to 0.58 μm if the refractive index 1.47 is assumed. This is between the usual sizes of PSC of types Ia and Ib. However, such narrow distribution is less expected for intermediate-sized and possibly double-phase particles.

Finally, the third minimum of $\chi_3^2$ is related to relatively small particles with a radius of about 0.25 μm. This size is typical for the low mode of PSC particle size distribution based on balloon measurements (Deshler *et al.*, 2000) and in satisfactory agreement with the low mode of lidar data (Jumelet *et al.*, 2009). It is worthy of note that the low-$\chi_3^2$ area in Figure 9 is narrow and horizontal. This means that best-fit value of $r_0$ is almost independent of $\sigma_D$ and can be found with a high level of accuracy. It is physically defined by the effectiveness of light scattering (the maximum of volume-equivalent scattering $S(r)/r^3$ and surface-area-equivalent scattering $S(r)/r^2$) for $r$ about half of a wavelength and fast change of polarization $p_0$ with increase of $r$, that is visible in Figure 8. For the evening of January 27, the median particle radius is $r_0$=0.234±0.012 μm if the refractive index 1.47 typical for PSC Ib particles is assumed. The distribution width $\sigma_D$ found by a minimum of $\chi_3^2$ is 1.44±0.03, which is also in agreement with both balloon and lidar data. It is worthy of note that this part of parametric line in Figure 8 is almost horizontal, so median radius is less dependent on color and defined basically by polarization, while distribution width is defined by color.

The observational characteristics of PSC on the morning of January 27 are less certain. However, three possible types of size distribution are also seen in the $\chi_3^2$-map built for the refractive index 1.47 in Figure 10. It should be added that balloon results (Deshler *et al.*, 2000) show the possibility of bimodal distribution (PSC Ia + Ib) with the median radii found here (about 0.25 and 1.5 μm). This can give the same observational effect; however, the color and polarization data is not enough to study the bimodal distribution numerically.

**7. Discussion and conclusion**

Polar stratospheric clouds play a significant role in stratosphere chemistry, by activating chlorine species that destroy ozone. Negative temperature trends in the stratosphere caused by greenhouse gases (Thompson *et al.*, 2012) can increase the rate of PSC formation and expand it to the lower latitudes. Stratospheric cooling and PSC visual frequency in northern Russia in the winter of 2019-2020 were maximal over the last dozens of years, and one can expect an increase in the PSC occurrence rate for the future.

The basic technique of ground-based measurements of PSC microphysical properties is lidar sounding. In this paper, the authors considered the possibility of PSC size measurements by using the simple technology of all-sky cameras. In this case, the data on PSC can be expanded by the use of a number of auroral cameras in high latitudes.

However, this analysis is more complicated than for NLC with small particles. Polarization and color measurements can be used separately for NLC analysis; two color channel data alone is enough to estimate the mean size. PSC particles are larger, the dependence of color and polarization on the size is not monotonous, and the procedure requires both types of information simultaneously. Even in this case, there are a number of possible distribution models with the same observational properties. An increase in the number of wavelengths and combined lidar and photometric measurements seem to be the way to solve this problem.

Another physical difficulty of PSC optical sounding is the presence of three types of clouds with different sizes, shapes, and refractive indices. Clouds of type II will cause strong fragmental changes in the brightness, color, and polarization of the twilight background. It can be the reason for possible large uncertainties of the measured parameters and the difficulty of size distribution retrieval. However, in several cases, if clouds are bright, expanded, and consist basically of one type of particles, the accuracy of performed analysis can be high. The radius estimation $r_0$=0.234±0.012 μm made for the evening of January 27, 2020 is an example of this. Particles with a diameter close to the wavelength scatter the light effectively. In this case, clouds of type Ib can be visually bright.

The procedure of cloud field separation suggested here can be used in future all-sky camera analysis of polar stratospheric and noctilucent clouds.

**Acknowledgments**

**Appendix. Validation of the cloud field separation method**

This work suggests a "spot analysis" method of cloud field separation against the twilight background effective for PSC. It differs from techniques used earlier for studying NLC, and here the authors apply the new method to these NLC measurements to compare the results. We use polarization data (Ugolnikov and Maslov, 2019) and color data (Ugolnikov *et al*., 2017) and then find the median radius of NLC particles assuming lognormal size distribution with $\sigma_D$=1.4 as it was done in those papers.

Figure A1 shows the dependency of the polarization of NLC on the evening of June 25, 2018, found by the "spot analysis" method compared with low-harmonics filtering followed by orthogonal clear sky reduction in (Ugolnikov and Maslov, 2019). We can see an overlapping agreement for both procedures in 1σ-interval. As a result, one can obtain close estimations of the median radius presented in Table A1. The same situation takes place for two other cases of NLC polarization measurements.

Color analysis of NLC is principally similar to the PSC case described in Chapter 5 of this paper. The only difference is that the color gradient of NLC by cos θ, $P_{G,R}$, is found instead of $W_{G,R}$, using the whole range of available scattering angles (Ugolnikov *et al*., 2017). This value is directly related to the median radius in the case of the Rayleigh-Gans approximation valid for small particles. Table A1 shows the values of $P_{G,R}$ and the median radius retrieved by "spot analysis" and by Fourier filtering used in the cited paper.

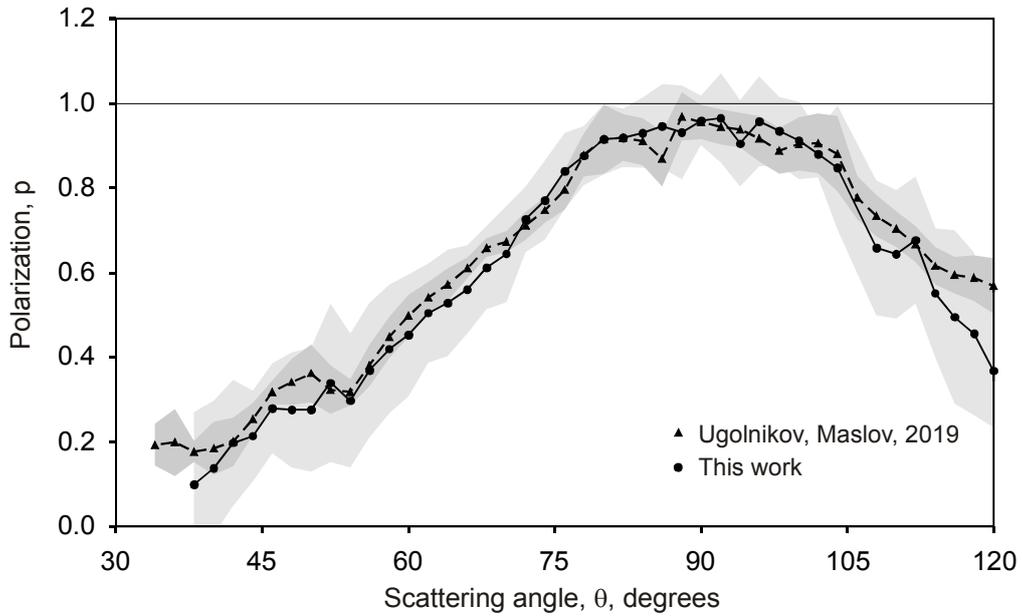

*Figure A1. Dependence of noctilucent clouds polarization on the scattering angle at June 25, 2018 retrieved by Ugolnikov and Maslov (2019) and by spot analysis in this work.*

The general similarity of results is worthy of note. For polarization measurements, the "spot analysis" method can lead to larger uncertainty in estimation; however, it is compensated by the confidence of background elimination from the measured data. It is especially important for PSC with a strong difference of color and polarization from the twilight background consisting mostly of Rayleigh scattering. Spot analysis has an advantage in accuracy over Fourier analysis for color data.

| Polarization (55.2°N, 37.5°E) | Ugolnikov and Maslov, 2019 | This work |
|---|---|---|
| July 5, 2015, $r_0$, μm | 0.027 ± 0.021 | 0.040 ± 0.035 |
| June 25, 2018, $r_0$, μm | 0.064 ± 0.009 | 0.061 ± 0.027 |
| June 27, 2018, $r_0$, μm | 0.074 ± 0.020 | 0.072 ± 0.028 |
| Color (68.0°N, 35.1°E), August 12, 2016 | Ugolnikov et al., 2017 | This work |
| $P_G$ | −0.063 ± 0.023 | −0.062 ± 0.013 |
| $r_0$, μm (G - B) | 0.027 ± 0.006 | 0.027 ± 0.003 |
| $P_R$ | −0.088 ± 0.038 | −0.063 ± 0.018 |
| $r_0$, μm (R - B) | 0.024 ± 0.005 | 0.021 ± 0.003 |

*Table A1. Color gradients and median radii of noctilucent clouds particles retrieved by spot analysis method of this work compared with existing techniques.*

**References**

Arctowskiy, H., 1902. Nuages lumineux et nuages irises, Ciel et terre, 1 Mar., 17-21.

Browell, E.V., Butler, C.F., Ismail, S., Robinette, P.A., Carter, A.F., Higdon, N.S., Toon, O.B., Schoeberl, M.R., and Tuck, A.F., 1990. Airborne lidar observations in the wintertime Arctic stratosphere: Polar stratospheric clouds, Geophys. Res. Lett., 17, 385.

Carslaw, K.S., Luo, B.P., Clegg, S.L., Peter, T., Brimblecombe, P., Crutzen, P. J., 1994. Stratospheric aerosol growth and $HNO_3$ gas phase depletion from coupled $HNO_3$ and water uptake by liquid particles. Geophysical Research Letters 21, 2479–2482.

Crutzen, P. J., Arnold, F., 1986. Nitric acid cloud formation in the cold Antarctic stratosphere: A major cause for the springtime "ozone hole". Nature, 324, 651.

Deshler, T., Nardi, B., Adriani, A., Cairo, F., Hansen, G., Fierli, F., Hauchecorne, A., Pulvirenti, L., 2000. Determining the index of refraction of polar stratospheric clouds above Andoya (69°N) by combining size-resolved concentration and optical scattering measurements. J. Geophys. Res., 105, D33943.

EOS MLS Science Team, 2011. MLS/Aura Level 2 temperature, version 003. Greenbelt, MD, USA: NASA Goddard Earth Science Data and Information Services Center (GES DISC). Accessed 01 May 2020 at) < https://acdisc.gesdisc.eosdis.nasa.gov/data/Aura_MLS_Level2/ML2T.004>.

Jumelet, J., Bekki, S., David, C., Keckhut, P., Baumgarten, G., 2009. Size distribution time series of a polar stratospheric cloud observed above Arctic Lidar Observatory for Middle Atmosphere Research (ALOMAR) (69°N) and analyzed from multiwavelength lidar measurements during winter 2005. J. Geophys. Res., 114, D02202.

Kassner, C., 1895. Irisirende wolken. Meteorologische Zeitschrift. 12, 379-382.

Molina, M.J., Zhang, R., Wooldridge, P.J., McMahon, J.R., Kim, J.E., Chang, H.Y., Beyer, K.D., 1993. Physical Chemistry of the $H_2SO_4/HNO_3/H_2O$ system: implications for polar stratospheric clouds. Science, 261, 1418.

Noel, V., Hertzog, A., Chepfer, H., Winker, D.M., 2008. Polar stratospheric clouds over Antarctica from the CALIPSO spaceborne lidar. J. Geophys. Res., 113, D02205.

Poole, L.R., McCormick, M.P., 1988. Airborne lidar observations of arctic polar stratospheric clouds: Indications of two distinct growth stages. Geophys. Res. Lett., 15, 21.

Rosen, J.M., 1971. The boiling point of stratospheric aerosols. Journal of Applied Meteorology, 10, 1044–1046.

Solomon, S., Garcia, R.R., Rowland, F.S., Wuebbles, D.J., 1986. On the depletion of Antarctic ozone. Nature 321, 755–758.

Solomon, S., 1990. Progress towards a quantitative understanding of Antarctic ozone depletion. Nature 347, 347–354.

Tabazadeh, A., Turco, R.P., Drdla, K., Jacobson, M.Z., 1994. A study of type I polar stratospheric cloud formation. Geophysical Research Letters, 21, 1619–1622.

Thompson, D.W.J., Seidel, D.J., Randel, W.J., Zou, Ch.-Z., Butler, A.H., Mears, C., Osso, A., Long, C., Lin, R., 2012. The mystery of recent stratospheric temperature trends. Nature, 491, 692.

Toon, O.B., Hamill, P., Turco, R.P., Pinto, J.P., 1986. Condensation of $HNO_3$ and HCl in the winter polar stratospheres. Geophys. Res. Lett, 13, 1284.

Ugolnikov, O.S., Kozelov, B.V., 2016. Mesosphere study by wide-field twilight polarization measurements: first results beyond the Polar Circle. Cosmic Research, 54, 279.

Ugolnikov, O.S., Maslov, I.A., Kozelov, B.V., Dlugach, Zh.M., 2016. Noctilucent cloud polarimetry: Twilight measurements in a wide range of scattering angles. Planetary and Space Science, 125, 105.

Ugolnikov, O.S., Galkin, A.A., Pilgaev, S.V., Roldugin A.V., 2017. Noctilucent cloud particle size determination based on multi-wavelength all-sky analysis. Planetary and Space Science, 146, 10.

Ugolnikov, O.S., Maslov, I.A., 2019. Polarization analysis and probable origin of bright noctilucent clouds with large particles in June 2018. Planetary and Space Science, 179, 104713.

Voigt, C., *et al.*, 2000. Nitric acid trihydrate (NAT) in polar stratospheric clouds, Science, 290, 1756.